\documentclass[%
 reprint,
 amsmath,amssymb,
 aps,
]{revtex4-2}

\usepackage{colortbl}
\usepackage{multirow}
\usepackage{graphicx}
\usepackage{amsmath}
\usepackage{amssymb}
\usepackage{amsthm}
\usepackage{amsfonts}
\usepackage{braket}
\usepackage{slashed}
\usepackage[T1]{fontenc}
\usepackage{mathtools}
\usepackage[utf8]{inputenc}
\usepackage[dvipsnames]{xcolor}
\usepackage{bm}
\usepackage{mathrsfs}
\usepackage{fontawesome}
\usepackage{tabularx}
\usepackage{booktabs}

\usepackage{caption}
\usepackage{subcaption}
\usepackage{hyperref}
\hypersetup{%
    colorlinks  = true,
    linkcolor   = [rgb]{0,0.3,0.7},
    citecolor   = [rgb]{0,0.3,0.7},
    urlcolor    = [rgb]{0,0.3,0.7},
    linktocpage = true
}

\begin{document}


\title{Tripartite entanglement of oscillating and decohering neutrinos}
\author{Subhashish Banerjee}
\affiliation{Departement of Physics, Indian Institute of Technology Jodhpur, India}

\author{Heinrich P\"as}
\affiliation{Department of Physics, Technische Universit\"at Dortmund, Germany}
\author{Erika Rani}
\altaffiliation[Also at ]{Department of Physics\, Universitas Islam Negeri Maulana Malik Ibrahim Malang, Indonesia}
\email{e.rani@fis.uin-malang.ac.id}
\affiliation{Department of Physics, Technische Universit\"at Dortmund, Germany}

\email{subhashish@iitj.ac.in}

\begin{abstract}
We study entanglement measures for oscillating neutrinos in the wave-packet formalism and confirm that an oscillating neutrino exhibits genuine tripartite entanglement and can be characterized as a member of the W class of states. We also show that this feature survives in the decoherence limit, and discuss the effect of CP phases on the entanglement. These findings provide new insights into the quantum nature of neutrino oscillations and their potential role in quantum information science, especially for neutrinos propagating large distances such as the  astrophysical neutrinos observed at neutrino telescopes.

\end{abstract}

\maketitle


\onecolumngrid
\section{Introduction}
In recent years, methods and techniques from quantum information science have become increasingly important for problems in quantum field
theory, particle physics and quantum gravity. Among the particles of the Standard Model, neutrinos are particularly interesting in this context, since their weak interactions qualify them as excellent quantum probes, allowing the genuine quantum phenomenon of neutrino oscillations to occur on scales as large as e.g. the Earth's diameter in the case of atmospheric neutrino oscillations. Above all, astrophysical neutrinos detected at neutrino telescopes such as IceCube or KM3Net may probe quantum information concepts and the quantum-to-classical transition on extragalactic scales and energies well above the 100 PeV range
\cite{IceCube:2014stg,KM3NeT:2025npi}.

 Entanglement constitutes arguably the most characteristic feature of quantum mechanics, and decoherence, i.e. the loss of coherence of a quantum system due to interactions with an environment, is generally considered as one if not the defining aspect of the quantum-to-classical transition. To establish a consistent framework for a quantum information theory analysis of these phenomena we thus study entanglement measures for oscillating and decohering neutrinos.
 Although entanglement is typically understood as a multi-particle phenomenon, single-particle phenomena can exhibit mode entanglement, also known as occupation number entanglement
  \cite{vanEnk:2005bhd,Cunha:2019jex}, and a series of previous works has shown
 that this is indeed the case for neutrino
 oscillations \cite{Blasone:2007vw,Blasone:2010ta,Blasone:2013zaa,Banerjee:2015mha,Blasone:2015lya,Alok:2014gya,Bittencourt:2018ywl,Dixit:2018kev,Song:2018bma,Dixit:2018gjc, Naikoo:2019eec,Jha:2020dav,Li:2022mus,Bittencourt:2022tcl,Banerjee:2024lih,Bouri:2024kcl}. 
These papers commonly use plane waves to study quantum entanglement measures, and thus assume an idealized scenario where neutrinos possess definite momentum and energy, and/or confine themselves to bipartite entanglement. However, 
a realistic description of the entanglement properties of oscillating neutrinos requires a complete 3-flavor analysis with multipartite entanglement in the wave packet formalism, since
neutrinos
are produced and detected as wave packets with a finite spread in momentum. Thus, the wave packet approach allows for a more precise description of neutrino-propagation and oscillation. Moreover, it naturally incorporates decoherence effects arising at
large $L/E$, where $L$ is the oscillation baseline and $E$ denotes the neutrino beam's energy, as the finite size of the wave packet causes a loss of coherence over long distances, which is essential for understanding neutrino behavior over astronomical scales and in low-energy experiments. 
We thus build up and extend on previous works  
that have studied genuine tripartite entanglement in a plane wave approximation \cite{Jha:2020dav} and others that discussed entanglement of oscillating neutrinos in a wave-packet approach, albeit confining the discussion to bipartite entanglement measures \cite{Blasone:2015lya} (compare also
\cite{Banerjee:2015mha,Li:2022mus}). 
These generalizations entail several subtleties that make the study of genuine tripartite entanglement for mixed states non-trivial, such as that relations between different entanglement measures hold in the simpler cases of bipartite entanglement or pure states but not in the general case discussed here.

This paper is organized as follows: In section II, we develop the formalism of three-flavor neutrino oscillations in the wave-packet formalism,
described as a three-qubit system in the occupation number basis. In section III, we introduce the relevant bipartite entanglement measures, i.e.
tangle, concurrence and negativity that characterize the distribution of entanglement between the three neutrino flavors, as well as the residues of these measures that quantify three-particle entanglement. In section IV, we study the behavior of these entanglement measures as a function of energy and oscillation baseline, and discuss the decoherence limit and the effects of CP phases. We summarize and draw conclusions in section IV.

\section{Neutrino oscillations as a two-qubit 
system in the wave-packet approach}

In the wave-packet formalism the neutrino flavor state $ \left|\nu_\alpha(x,t)\right>$, $\alpha={e, \mu, \tau}$
is given by the weighted superposition of mass eigenstates $\left|\nu_s\right>$, $s=1, 2, 3$, 
\begin{equation}
\label{2.1}
    \left|\nu_\alpha(x,t)\right>= \sum_s V_{\alpha s}^*(\theta)\, \psi_s (x,t)\left|\nu_s\right>.
\end{equation}
Here $V$ denotes the
Pontecorvo-Maki-Nakagawa-Sakata (PMNS) matrix
\begin{equation}
    V=\begin{pmatrix}
        c_{12} c_{13}& s_{12}c_{13}&s_{13}e^{i\delta_{CP}}\\ -s_{12}c_{23}-c_{12}s_{13}s_{23}e^{i\delta_{CP}}&c_{12}c_{23}-s_{12}s_{13}s_{23}e^{i\delta_{i\delta_{CP}}}&c_{13}s_{23}\\ s_{12}s_{23}-c_{12}s_{13}c_{23}e^{i\delta_{CP}}&-c_{12}s_{23}-s_{12}s_{13}c_{23}e^{i\delta_{CP}}&c_{13}c_{23}
    \end{pmatrix},
\end{equation}
where $c_{ij}\equiv \cos\theta_{ij}$ and $s_{ij}\equiv \sin\theta_{ij}$, with $i,j=1,2,3$, 
and $\delta_{CP}$ denoting the Dirac CP phase.
The two Majorana CP phases that do not affect neutrino oscillations are set to zero for convenience.

The wave-packet $\psi_s(x,t)$ has the following form:
\begin{eqnarray}
    \psi_s(x,t) = \frac{1}{\sqrt{2\pi}}\int \mathrm{dp} \, \psi_s(p) \  e^{i(px-E_s(p)t)},
\end{eqnarray}
where $E_s(p)= \sqrt{p_s^2+m^2_s}$ is the energy of the mass eigenstate $s$. The wave function $\psi_s(p)$ results as:
\begin{eqnarray}
\label{2.2}
    \psi_s(p) = (2\pi\sigma_P^2)^{1/4} \exp{\left[-\frac{(p-p_s)^2}{4\sigma_P^2}\right]},
\end{eqnarray}
where $p_s$ is the mean momentum and $\sigma_p$ denotes the momentum width of the wave packet. Assuming the wave function reaches its peak around $\sigma_p \ll E^2_s(p_s)/m_s$, we can estimate the energy $E_s(p)$ by $E_s(p)\simeq E_s+ v_s(p-p_s)$ where $v_s=\frac{p_s}{E_s}$ is the group velocity of wave packet, and $s=1,2,3$. After a Gaussian integration over $p$ in 
eq.~(\ref{2.1}), the neutrino flavor state develops as
\begin{eqnarray}
\label{2.3}
    \left|\nu_\alpha(x,t)\right> &=& (2\pi\sigma_x^2)^{-1/4}\sum_s
V_{\alpha s}^* 
\times \exp{\left[ip_s x-iE_s t-\frac{(x-v_st)^2}{4\sigma_x^2}\right]}\left|\nu_s\right>,
\end{eqnarray}
where the mass eigenstates $\left|\nu_s\right>$ fulfill the orthonormality relationship $\left<\nu_s|\nu_r\right>=\delta_{sr}$, and  $\sigma_x$ is the width of wave packets in coordinate space related to the width in momentum space $\sigma_p$ as $\sigma_x=(2\sigma_p)^{-1}$.
\par

To calculate the oscillation probability, we adopt the following approximations \cite{Blasone:2015lya}:
\begin{equation}
\label{2.4}
    E_s \cong E,\hspace{1cm} p_s\cong E-\frac{m_s^2}{2E},\hspace{1cm} v_s\cong 1-\frac{m_s^2}{2E_s^2}
\end{equation}
where $E$ is the neutrino energy in the limit of vanishing neutrino masses.
The density matrix of the pure state follows as:
\begin{eqnarray}
\label{2.5}
    \rho^{(\alpha)}(x,t)= \left|\nu_\alpha(x,t)\right>\left<\nu_\alpha(x,t)\right|,
\end{eqnarray}
where the index $\alpha$ specifies the original neutrino flavor at the source.
Using the approximation in (\ref{2.4}), we can obtain a space-time picture of neutrino oscillation via the density matrix which is calculated by inserting eq.(\ref{2.3}) into (\ref{2.5}), and performing a Gaussian integration over time. Finally, the density matrix is obtained as \cite{Blasone:2015lya}:
\begin{eqnarray}
\label{2.6}
    \rho^{(\alpha)}(x) &=& \sum_{sr} V_{\alpha s}V^*_{\alpha r} \varphi_{sr}(x)\left|\nu_s\right>\left<\nu_r\right|,\\
    \label{2.6a}
    \varphi_{sr}&\equiv& \exp{\left[-i\frac{\Delta m^2_{sr}x}{2E}-\left(\frac{\Delta m^2_{sr} x}{4\sqrt{2}E^2\sigma_x}\right)^2 \right]} ,
\end{eqnarray}
with $\Delta m^2_{sr}= m_s^2-m_r^2$. 

The density matrix in Eq. (\ref{2.5})  describes a mixed state rather than a pure state when the wave-packet treatment is taken into account. Although the neutrino is initially produced in a pure flavor state, the wave-packet treatment introduces an intrinsic spread in momentum and leads to slightly different group velocities for the mass eigenstates. As the neutrino propagates, these wave packets gradually separate in space. The loss of spatial overlap between different mass-eigenstate wave packets suppresses the off-diagonal elements of the density matrix via the Gaussian damping factor in Eq. (\ref{2.6a}) and induces decoherence.

To analyze the mode entanglement of an oscillating neutrino, we 
represent the neutrino flavor states in the occupation number basis as follows:
\begin{eqnarray}
\label{2.3aa}
    \left|\nu_e\right> &=& \left|1\right>_e\otimes\left|0\right>_\mu\otimes\left|0\right>_\tau\equiv\left|100\right>_e\\
     \left|\nu_\mu\right> &=& \left|0\right>_e\otimes\left|1\right>_\mu\otimes\left|0\right>_\tau\equiv\left|010\right>_\mu\\
     \label{2.3ab}
\left|\nu_\tau\right> &=& \left|0\right>_e\otimes\left|0\right>_\mu\otimes\left|1\right>_\tau\equiv\left|001\right>_\tau.
\label{2.3ac}
\end{eqnarray}
Furthermore, with 
$\left|\nu_s\right>=\sum_\alpha V_{\alpha s}\left|\nu_\alpha\right>$,
$\alpha=e,\mu,\tau$, the density matrix in Eq. (\ref{2.6}) is given by
\begin{eqnarray}
\label{2.7}
    \rho^{(\alpha)}(x) = \sum_{\beta\gamma} F_{\beta\gamma}^\alpha(x) \left|\nu_\beta\right>\left<\nu_\gamma\right|,
\end{eqnarray}
and
\begin{eqnarray}
    F_{\beta\gamma}^{\alpha}(x)= \sum_{sr} V^*_{\alpha s}V_{\alpha r}\varphi_{sr}V_{\beta s} V_{\gamma r}^*,
\end{eqnarray}
where the indices are $s,r= 1,2,3$, and $\beta,\gamma = e,\mu,\tau$. \par

\section{Entanglement Measures in Neutrino Oscillations  }

\begin{figure}[h!]
    \centering
    \includegraphics[width=.8\linewidth]{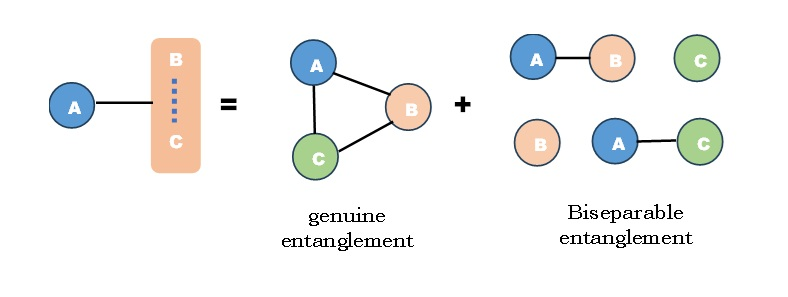}
    \caption{Possible realizations of three-qubit entanglement: Apart from product states, three qubits
    can be in a biseparable state, i.e. a product of a single state and a pair of qubits with bipartite entanglement, or can exhibit genuine tripartite entanglement.}
    \label{skema}
\end{figure}

The entanglement among three qubits A, B and C can be classified by assigning the quantum state to one of six classes that cannot be transformed into each other by stochastic local operations and 
classical communcation (SLOCC) \cite{Dur:2000zz}. These classes are the A$\otimes$B$\otimes$C product states, the A$\otimes$BC, B$\otimes$AC and C$\otimes$AB biseparable states, i.e. product states of an individual qubit and a bipartite entangled pair of qubits, and the genuine tripartite entanglement GHZ and W classes of states with representatives $|GHZ\rangle = 1/\sqrt{2} \left(|000\rangle + |111\rangle \right)$ and
$|W\rangle = 1/\sqrt{3} \left(|001\rangle + |010\rangle + |001\rangle \right)$, see Fig.~\ref{skema} (compare also \cite{Cunha:2019jex}).  
\par

Multipartite entanglement, or more concretely, the extent and distribution of entanglement between various qubits can be 
described and quantified by various entanglement measures, see \cite{Horodecki:1998kf,Wootters:1997id,Vedral:1997qn} for bipartite entanglement, \cite{Coffman:1999jd}
for three-qubit entanglement, and \cite{Horodecki:2009zz} for a review. In this paper, we will analyze the entanglement of oscillating neutrinos by employing the entanglement measures tangle, concurrence, negativity, and the residual three-$\pi_{e\mu\tau}$
negativity.

We quantify the tripartite entanglement by rewriting the density matrix Eq. (\ref{2.7}) in matrix form within the standard basis $\left|ijk\right>$ with $i,j,k=\{0,1\}$ as
\begin{eqnarray}
\label{2.8}
    \mathcal{\rho}_{e\mu\tau}^{(\alpha)}(x)= \begin{pmatrix}
        0&0&0&0&0&0&0&0\\
        0&0&0&0&0&0&0&0\\
        0&0&0&0&0&0&0&0\\
        0&0&0&F_{ee}^\alpha(x)&0&F_{e\mu}^\alpha(x)& F_{e\tau}^\alpha(x)&0\\
        0&0&0 &0&0&0&0&0\\
         0&0&0& F_{\mu e}^\alpha(x)&0&F_{\mu\mu}^\alpha(x)&F_{\mu\tau}^\alpha(x)&0\\
          0&0&0&F_{\tau e}^\alpha(x)&0&F_{\tau\mu}^\alpha(x)&F_{\tau\tau}^\alpha(x)&0\\
           0&0&0&0&0&0&0&0\\
    \end{pmatrix}.
\end{eqnarray}

In the other words, the matrix element of the density matrix in Eq. (\ref{2.7}) is expressed as 

\begin{eqnarray}
\label{2.8a}
\left(\rho^{(\alpha)}_{e\mu\tau}\right)_{i'j'k',ijk}(x)=\left<i'\right|\left<j'\right|\left<k'\right|\rho_{e\mu\tau}^{(\alpha)}(x)\left|i\right>\left|j\right>\left|k\right>,
\end{eqnarray}
where the indices $i,j,k,i',j',k'$ can adopt the values 0 or 1.

\subsection{Concurrence and Tangle}
In the following, we introduce and discuss the entanglement measures studied in this work, i.e., concurrence and tangle, negativity, the CKW inequality, and the genuine tripartite entanglement.
Concurrence has been introduced by Wooters in 1998 and quantifies how much a 2-qubit state is affected by a spin flip transformation. For example a maximal entangled state like the 2-spin singlet that remains unchanged up to an overall sign has concurrence 1, while a product state
that gets transformed into an orthogonal state has concurrence zero 
\cite{Wootters:1997id}. More concretely, 
a general two-qubit state $\psi$ can be written as
\begin{equation}
    \left|\psi\right>_{AB}= A\left|10\right>+ B\left|01\right>
\end{equation}
where $|A|^2+|B|^2=1$. Wootters et al. \cite{Wootters:1997id,Coffman:1999jd} define the "spin-flipped" density operator of the two-qubit system, $\tilde{\rho}_{AB}=(\sigma_y\otimes\sigma_y)\rho^*_{AB}(\sigma_y\otimes\sigma_y)$, where $\sigma_y = \begin{pmatrix}
    0&-i\\i&0
\end{pmatrix}$ is a Pauli matrix, and $\rho^*_{AB}$ is the complex conjugation of $\rho_{AB}$. Both $\rho_{AB}$ and $\tilde\rho_{AB}$ are positive operators. 

The concurrence for a two-qubit state can be written in the form \cite{Wootters:1997id}

\begin{equation}
\label{Con1}
    C_{AB} = \left[\mathrm{max}\{0,\lambda_1-\lambda_2-\lambda_3-\lambda_4\}\right],
\end{equation}
where $\lambda_1,\lambda_2,\lambda_3,\lambda_4$ are the square roots of the eigenvalues obtained from the product of the non-Hermitian matrix $\rho_{A(BC)}\tilde{\rho}_{A(BC)}$ in descending order.
For a pure state, the concurrence reduces to $C_{AB} = 2 \sqrt{\rm{det}\rho_A}$, where $\rho_A = \rm{Tr}_B (\rho_{AB})$. This definition has been used in the plane-wave discussion of tripartite entanglement in \cite{Jha:2020dav} but it is not applicable in the wave-packet approach pursued in this work.
As we are dealing with a mixed state where the wave packets propagate and decoherence suppresses the off-diagonal terms, the concurrence is defined using the general expression given in Eq.~(\ref{Con1}) \cite{Coffman:1999jd}.  
For an oscillating neutrino produced in the flavor $\alpha$ at the source, the concurrence of the bipartite system of flavors $\beta$, $\gamma$ is given by
\begin{eqnarray}
\label{conplot}
    C_{\alpha}^{(\beta\gamma)}= 2|F^{(\alpha)}_{\beta\gamma}(x)|.
\end{eqnarray}

Bipartite entanglement in a three-qubit neutrino oscillation system can be analyzed by studying e.g. the entanglement between the
$e$ flavor adopted as the reference and the remaining $\mu\tau$ flavors, where the latter flavors are treated as a single object. Likewise, when the entanglement of $\mu$ or $\tau$ flavors is studied, the $e\tau$ and $e\mu$ states are considered as single quantum objects, respectively. For example, the squared concurrence of a multipartite neutrino system, for the bipartition $e|(\mu\tau)$, is given by:
\begin{eqnarray}
\label{2.9z}
   \left( C_{\alpha}^{e(\mu\tau)}\right)^2 = 4( |F_{e\mu}^{\alpha}|+|F_{e\tau}^{\alpha}|)^2
\end{eqnarray}
Closely related to the concurrence is the tangle, which is defined as the square of the concurrence, $\tau_{AB}= C^2_{AB}$. For the three-flavor neutrino system, the tangle for the bipartition $e|(\mu\tau)$ is given by $\tau_{e(\mu\tau)}(x)=C^2_{e(\mu\tau)}(x) = 4( |F_{e\mu}^{\alpha}|+|F_{e\tau}^{\alpha}|)^2$ which shows that entanglement between the $\mathit{e}$-flavor and the pair $(\mu\tau)$ is determined by the off-diagonal coherence of the density matrix.

\subsection{Negativity}
To define the entanglement measure negativity, we employ the
the positive partial transpose (PPT) criterion. The partial transpose applied on the subsystem $e,\mu$ or $\tau$ in Eq.~(\ref{2.8a}) is given by \cite{Jha:2020dav}
\begin{eqnarray}
\left(\rho^{(\alpha)T_{e}}_{e\mu\tau}\right)_{i'j'k',ijk}&=&\left(\rho^{(\alpha)}_{e\mu\tau}\right)_{ij'k',i'jk} \nonumber \\ \left(\rho^{(\alpha)T_{\mu}}_{e\mu\tau}\right)_{i'j'k',ijk}&=&\left(\rho^{(\alpha)}_{e\mu\tau}\right)_{i'jk',ij'k}\nonumber \\
\left(\rho^{(\alpha)T_{\tau}}_{e\mu\tau}\right)_{i'j'k',ijk}&=&\left(\rho^{(\alpha)}_{e\mu\tau}\right)_{i'j'k,ijk'},  
\end{eqnarray}
respectively.\par

The PPT criterion, first introduced by Peres 
\cite{Peres:1996dw,Horodecki:2009zz}, provides a test for entanglement based on the fact that 
a density matrix $\rho$ of a pure or mixed separable state 
can have only positive eigenvalues.Taking the partial transpose of Eq. (\ref{2.8a}) with flavor $e$ chosen as the reference for the bipartition $e|(\mu\tau)$, yields the following eigenvalues of $\rho^{(\alpha)T_e}_{e(\mu\tau)}(x)$:
\begin{eqnarray}
    \mu_1 &=& |F_{e\mu}^{\alpha}(x)|\nonumber\\
    \mu_2&=& -|F_{e\mu}^{\alpha}(x)|\nonumber\\
    \mu_3&=& |F_{e\tau}^{\alpha}(x)|\nonumber\\
    \mu_4&=& -|F_{e\tau}^{\alpha}(x)|\nonumber\\
       \mu_{5,6} &=& \frac{F_{\mu\mu}^{\alpha}(x)+F_{\tau\tau}^{\alpha}(x)}{2}\pm \frac{\sqrt{(F_{\mu\mu}^{\alpha}(x)F_{\tau\tau}^{\alpha}(x))^2+4 |F_{\mu\tau}^{\alpha}(x)|^2}}{2} \nonumber\\
    \mu_7&=&F_{ee}^{\alpha}(x)\nonumber\\
    \mu_8 &=&0.
\end{eqnarray}
As we obtain one negative eigenvalue, $\rho^{(\alpha)T_e}_{e(\mu\tau)}(x)$ is not represented by a positive operator, and consequently the flavor modes $e$, $\mu$, and $\tau$ are entangled. The negativity of a quantum state quantifies its entanglement by measuring how
strongly the PPT criterion is violated.

Using the trace norm of $\rho^{T_a}$ , 
\begin{equation}
    ||\rho^{T_a}||= Tr\sqrt{\rho^{T_a}\rho^{T_{a}^\dagger}} = 1+2|\sum_i\mu_i|,
\end{equation}
the entanglement negativity is defined by \cite{Ou:2007ysv}:
\begin{equation}
    N(\rho)\equiv\frac{||\rho^{T_a}||-1}{2}
\end{equation}
where 
$\mu_i<0$ are the non-positive eigenvalues of $\rho^{T_a}$. The negativity of the bipartition of the tripartite state $e|(\mu\tau)$ results, as
\begin{eqnarray}
\label{2.9}
  N_{e(\mu\tau)}^{T_e}(x)&\equiv& N_{e(\mu\tau)}(x)\nonumber\\
  &=& |F_{e\mu}^\alpha(x)|+|F_{e\tau}^\alpha(x)|. 
\end{eqnarray}
By comparing Eqs. (\ref{2.9z}) and (\ref{2.9}), one obtains the relation:
\begin{eqnarray}
\label{2.9a}
    \tau_{e(\mu\tau)}(x)= C^2_{e(\mu\tau)}(x) = 4 N_{e(\mu\tau)}^2(x) = 4 \left(|F_{e\mu}^\alpha(x)|+|F_{e\tau}^\alpha(x)|\right)^2.
\end{eqnarray}
Note that Eq.~(\ref{2.9a}) differs from the analogous expression in \cite{Jha:2020dav} by a numerical factor of 4 in front of the negativity that seems to be omitted 
erroneously in \cite{Jha:2020dav}. Note further that the analogous expression to Eq.~(\ref{2.9a}) for bipartite (i.e. 2-flavor) states
derived in \cite{Jha:2020dav} for pure states does not exist in the case
of mixed states discussed here. One consequence of this finding is that
the CKW inequality for negativities is not saturated as it is for tangles, see below and \cite{Ou:2007ysv}. 

\subsection{Monogamy of Entanglement and the CKW inequality} 
A fundamental principle in quantum information science is the monogamy of entanglement that describes how 
two qubits A and B that are maximally entangled cannot be entangled with a third qubit C. If conversely the 
two qubits are only partially entangled, they can also be
partially entangled with the third qubit C.
The concept of monogamy can be quantified using monogamy inequalities such as the  Coffman-Kundu-Wootters (CKW) inequality \cite{Coffman:1999jd},
\begin{equation}
\label{1.1}
     \tau(\rho_{AB}) + \tau(\rho_{AC}) \leq \tau(\rho_{A(BC)}) 
\end{equation}
where $ \tau(\rho_{A(BC)})$ is a tangle quantifying the degree of entanglement between qubit A and the pair of qubits B and C considered as a single object, and  
$\tau(\rho_{AB})$, $ \tau(\rho_{AC})$ are the bipartite tangles between qubits A and B, or A and C, respectively. An
interesting consequence of the CKW inequalities is the residual entanglement in terms of tangle, $\tau_A = \tau_{A(BC)}-\tau_{AB}-\tau_{AC}$, that can quantify the entanglement in the
three-qubit system that cannot be described by the bipartite
entanglement specified by the tangles $\tau_{AB}$ and $\tau_{AC}$ \cite{Ou:2007ysv}.

Adopting the $e$ flavor as the reference qubit A,
the residual tangle for the electron neutrino follows as
\begin{eqnarray}
\label{2.11}
    \tau_{e} = \tau_{e(\mu\tau)}- \tau_{e\mu}-\tau_{e\tau}.
    \end{eqnarray}
Similarly, if one adopts $\mu$ and $\tau$ as the reference mode, the residual entanglements $\tau_\mu$ and $\tau_\tau$
results as
\begin{eqnarray}
   \tau_{\mu} = \tau_{\mu(e\tau)}- \tau_{\mu e}- \tau_{\mu\tau}\\
   \label{2.11a}
    \tau_{\tau} = \tau_{\tau(e\mu)}- \tau_{\tau e}-\tau_{\tau\mu},
\end{eqnarray}
respectively.

If the entanglement in the system can be fully described by the bipartite entanglement measures, the residual entanglement is zero. Thus, a non-zero residual entanglement indicates the existence of genuine tripartite correlations beyond simple bipartite contributions \cite{Coffman:1999jd}. 
However, for W-type states, such as the neutrino oscillating state, the residual tangle vanishes even though genuine multipartite entanglement is still present \cite{Dur:2000zz}. Although first formulated as a condition for tangles, the
CKW inequality can also be generalized 
for concurrence \cite{Kim:2008juv} and negativity \cite{Ou:2007ysv}. 

\subsection{Genuine Tripartite Entanglement}
\par
In general the entanglement of a tripartite quantum system
cannot be completely described in terms of bipartite entanglement measures but is 
rather distributed between all three flavor modes, as is the case for the GHZ and W classes of states.  

As has been discussed in Eqs. (\ref{2.11})-(\ref{2.11a}) and will be shown in the following in Fig.~\ref{ckw},
however, concurrence and tangle are no suitable measures to describe the genuine entanglement of W-class states.
Therefore, in the following, we thus consider the residual entanglement in terms of negativity. Analogous to the CKW inequality for tangles, there exists a CKW inequality in terms of negativity:
\begin{eqnarray}
\label{2.14}
    N_{e\mu}^2+N_{e\tau}^2 \leq N_{e(\mu\tau)}^2,
\end{eqnarray}
where $N_{e\mu}$ and $N_{e\tau}$ are the negativities of the mixed states $\rho_{e\mu}^\alpha = \mathrm{Tr}_\tau(\rho^\alpha_{e\mu\tau}(x))$ and $\rho_{e\tau}^\alpha = \mathrm{Tr}_\mu(\rho^\alpha_{e\mu\tau}(x))$, respectively  \cite{Ou:2007ysv}. 
Analogously, if one adopts the
$\mu$ and $\tau$ flavors as a reference, the corresponding
CKW inequalities are
\begin{eqnarray}
    \label{2.15}
    N_{\mu e}^2+N_{\mu\tau}^2 \leq N_{\mu(e\tau)}^2
\end{eqnarray}
and
\begin{eqnarray}
    \label{2.16}
    N_{\tau e}^2+N_{\tau\mu}^2 \leq N_{\tau(e\mu)}^2,
\end{eqnarray}
respectively.

By choosing the $e$-flavor mode as the reference, the squared product negativity of the entangled $e-\mu$ and $e-\tau$  are given by
\begin{eqnarray}
    N_{e\mu}^2 &=& \frac{\left(\sqrt{(F_{\tau\tau}^\alpha (x))^2+ 4|F_{e\mu}^\alpha(x)|^2}-F_{\tau\tau}^\alpha (x)\right)^2}{4},
\end{eqnarray}
and
\begin{eqnarray}
    N_{e\tau}^2 &=&\frac{\left(\sqrt{(F_{\mu\mu}^\alpha (x))^2+ 4|F_{e\tau}^\alpha(x)|^2}-F_{\mu\mu}^\alpha (x)\right)^2}{4}.
    \label{Netausquaredq}
\end{eqnarray}
Substituting these expressions into the CKW inequality in Eq.~(\ref{2.14}), we obtain the sum of $N_{e\mu}^2(x)$ and $N_{e\tau}^2(x)$ as
\begin{eqnarray}
\label{2.9b}
    N_{e\mu}^2(x)+N^2_{e\tau}(x) &=& \frac{(F_{\tau\tau}^\alpha (x))^2+(F_{\mu\mu}^\alpha (x))^2}{2} + |F_{e\mu}^\alpha(x)|^2+|F_{e\tau}^\alpha(x)|^2\nonumber\\
    &&-\frac{1}{2}\left[ F_{\tau\tau}^\alpha (x)\sqrt{(F_{\tau\tau}^\alpha (x))^2+ 4|F_{e\mu}^\alpha(x)|^2} 
    +  F_{\mu\mu}^\alpha (x)\sqrt{(F_{\mu\mu}^\alpha (x))^2+ 4|F_{e\tau}^\alpha(x)|^2}  \right].
\end{eqnarray}
From the last term in Eq.~(\ref{2.9b}), we have  $F_{\tau\tau}^\alpha(x)\sqrt{(F_{\tau\tau}^\alpha (x))^2+ 4|F_{e\mu}^\alpha(x)|^2} \geq (F_{\tau\tau}^\alpha (x))^2 $ and $F_{\mu\mu}^\alpha(x)\sqrt{(F_{\mu\mu}^\alpha (x))^2+ 4|F_{e\tau}^\alpha(x)|^2} \geq (F_{\mu\mu}^\alpha (x))^2 $. Thus, the last term is bigger than the first term. Comparing Eq. (\ref{2.9}) and Eq. (\ref{2.9b}), we find that the CKW inequality in terms of negativity for the neutrino-flavor system is satisfied.

The residual entanglement in terms of negativities for the $e$, $\mu$, and $\tau$ flavor adopted as references are given by $\pi_e= N_{e(\mu\tau)}^2-N_{e\mu}^2-N_{e\tau}^2$, $\pi_\mu= N_{\mu(e\tau)}^2-N_{\mu e}^2-N_{\mu\tau}^2$, and $\pi_\tau= N_{\tau(e\mu)}^2-N_{\tau e}^2-N_{\tau\mu}^2$, respectively. 

Finally, the three-$\pi$negativity $\pi_{e\mu\tau}$ \cite{Ou:2007ysv}, i.e., the average of $\pi_e$, $\pi_\mu$, and $\pi_\tau$ provides a 
suitable measure of genuine tripartite entanglement for oscillating neutrinos,
\begin{eqnarray}
\label{aa}
    \pi_{e\mu\tau} = \frac{1}{3}( \pi_e+ \pi_\mu+\pi_\tau),
\end{eqnarray}
One obtains
\begin{eqnarray}
    \pi_{e\mu\tau}&=& \frac{2}{3}\left( |F_{e\mu}^\alpha||F_{e\tau}^\alpha (x)|+|F_{e\mu}^\alpha||F_{\mu\tau}^\alpha (x)|+|F_{e\tau}^\alpha||F_{\mu\tau}^\alpha (x)|\right)- \frac{1}{3}\left( (F_{ee}^\alpha(x))^2+(F_{\mu\mu}^\alpha(x))^2+(F_{\tau\tau}^\alpha(x))^2  \right)\nonumber\\
    &&+\frac{1}{3}\left[F_{\tau\tau}^\alpha(x)\sqrt{(F_{\tau\tau}^\alpha (x))^2+ 4|F_{e\mu}^\alpha(x)|^2}\right.\nonumber\\
    &&\left.+F_{\mu\mu}^\alpha(x)\sqrt{(F_{\mu\mu}^\alpha (x))^2+ 4|F_{e\tau}^\alpha(x)|^2}+ F_{ee}^\alpha(x)\sqrt{(F_{ee}^\alpha (x))^2+ 4|F_{\mu\tau}^\alpha(x)|^2} \right]>0.
\end{eqnarray}
 As the three-tangle, the three-$\pi$ in Eq.(\ref{aa}) is a natural entanglement measure satisfying the conditions of local unitary, 
 $\pi_{e\mu\tau}\geq 0$ and invariance under permutations of the qubit modes; see Eqs.~(\ref{2.14})-(\ref{2.16}) and \cite{Vedral:1997qn,Ou:2007ysv}. In contrast to the three tangle though, the three-$\pi$-negativity $\pi_{e\mu\tau}$ is also applicable to describe genuine tripartite entanglement also  for W-class states.
 
\section{Results}
In the following we discuss the distribution of entanglement between the flavor states of an oscillating and decohering neutrino.
To do so, we analyze the development of entanglement measures, i.e., tangle, concurrence, negativity, and the residual three-$\pi_{e\mu\tau}$negativity as a function of energy and the oscillation baseline. For quantitative results,  we adopt the neutrino oscillation parameters reported by NuFIT 6.0 \cite{Esteban:2024eli} and  the Particle Data Group (PDG) \cite{ParticleDataGroup:2024cfk}:
\begin{eqnarray}
    \Delta m^2_{21} &=& 7.49 \times 10^{-5} \mathrm{eV^2},\nonumber\\
    \Delta m^2_{31} &=& 2.51 \times 10^{-3}\mathrm{eV^2}, \nonumber\\
    \sin^2\theta_{12}&=& 0.308,\nonumber\\ \sin^2\theta_{13}&=& 2.215\times 10^{-2},\nonumber\\
    \sin^2\theta_{23}&=& 0.470.
\end{eqnarray}
The Dirac CP phase $\delta$ is treated as a free parameter
and either varied or assumed to be zero.
\par
We are particularly interested in the entanglement of neutrino 
oscillations over large baselines studied at neutrino telescopes, such as atmospheric neutrino oscillations with energies ranging from 0.1 GeV to 10 TeV and baselines from $10^6$~m (down-going) to 
$10^7$~m (up-going), and astrophysical or extra-galactic neutrinos
in the energy range of 60~TeV up to 100~PeV. In the latter case,
it is difficult to identify individual sources to obtain
concrete baselines, but for instance, the candidate source blazar TXS 0506+056 has a distance to Earth of $10^{25}$~m
(for a review, see e.g. \cite{Halzen:2021ynx}).
\par
The effects of decoherence due to wave packet separation have not been observed yet, and there exist only loose constraints with a spread of 13 orders of magnitude on the size of neutrino wave packages from reactor neutrino data, see e.g.
\cite{DayaBay:2016ouy,
deGouvea:2021uvg,
deGouvea:2024syg,
Smolsky:2024uby}
and for theoretical analyses \cite{Giunti:1997wq,Akhmedov:2019iyt,Akhmedov:2022bjs}.
For a better comparison with \cite{Blasone:2015lya}, we adopt the 
value used there,
$\sigma_p= 1$~GeV.

\begin{figure}[h!]
    \centering
    \begin{subfigure}[b]{0.6\textwidth}
        \includegraphics[width=\linewidth]{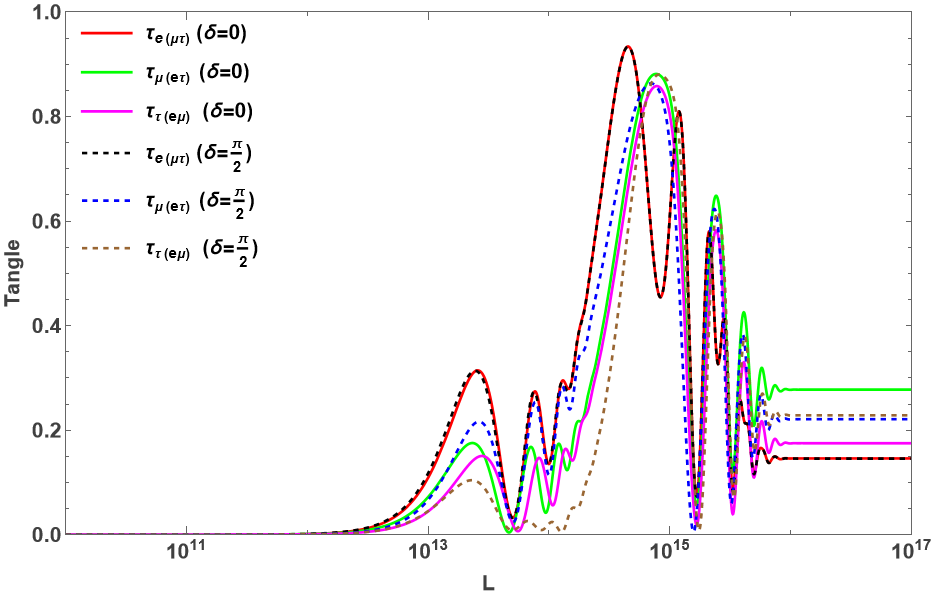}
        \caption{e-channel}
        \label{Tangs}
    \end{subfigure}
    \begin{subfigure}[b]{0.6\textwidth}
        \includegraphics[width=\linewidth]{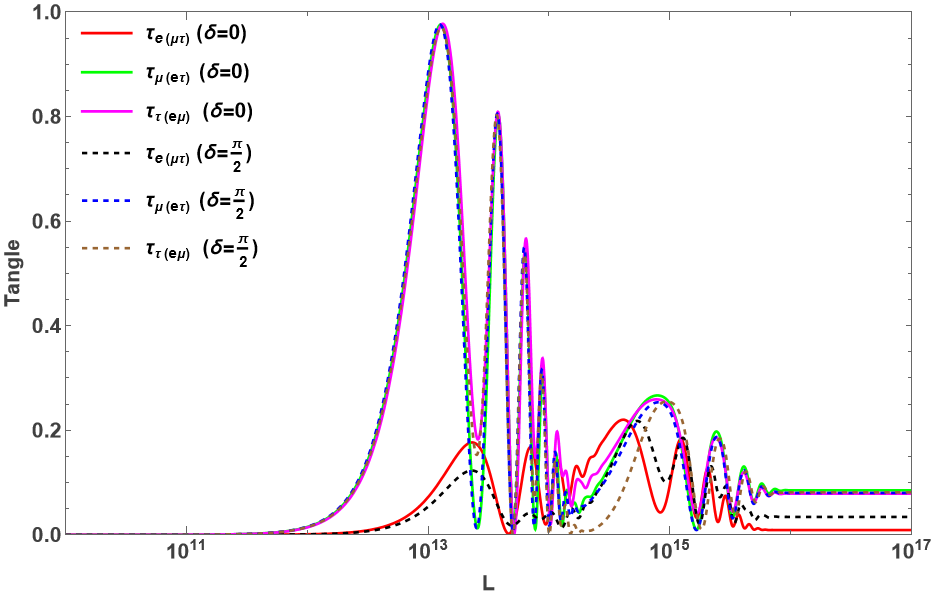}
        \caption{$\mu$-channel}
        \label{Tangss}
    \end{subfigure}
    \hfill
    \begin{subfigure}[b]{0.6\textwidth}
        \includegraphics[width=\linewidth]{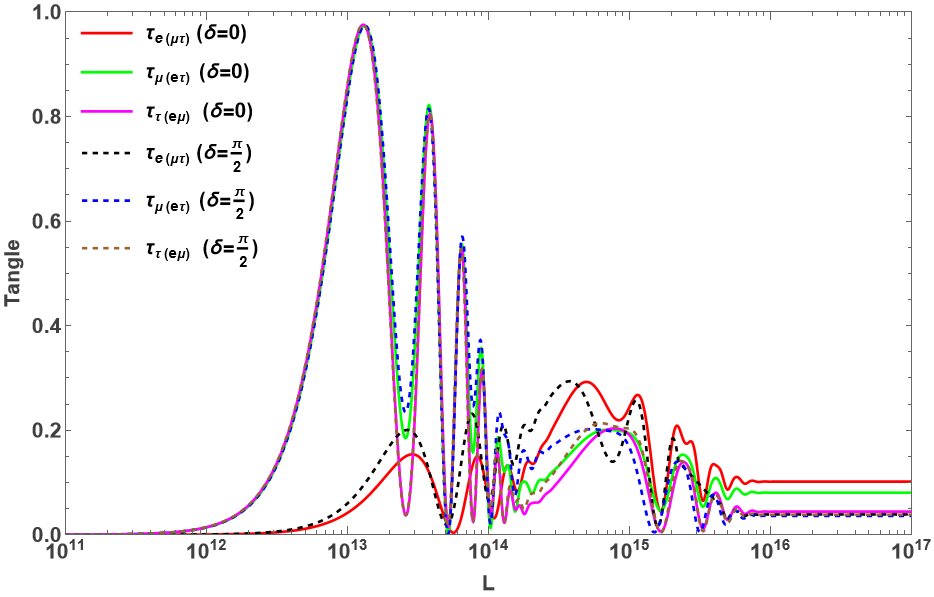}
        \caption{$\tau$ channel}
        \label{Tang3}
    \end{subfigure}
    \caption{Distribution of bipartite entanglement (tangle) between an individual flavor and the pair of remaining flavors for oscillating neutrinos with initial states: (a) $\nu_e$, (b) $\nu_\mu$, and (c) $\nu_\tau$ as a function of distance $L$ (in meters). The energy $E$ is set to 10~GeV and the CP-violating phase $\delta$ is chosen $\delta = 0, \frac{\pi}{2}$ in solid and dashed lines, respectively.}
    \label{Tang 10GeV}
\end{figure}

In Fig.~\ref{Tang 10GeV} we show the distribution of entanglement, quantified by the tangle as a function of baseline for three different initial neutrino flavor state : (a) electron, (b) muon, and (c) tau channels. In each channel, the tangle describes the entanglement between the initially prepared flavor and the pair formed by the remaining two flavors. At short distances, the initial tangle is zero, since the state is described by a single flavor with negligible entanglement. As $L$ increases, oscillation effects generate quantum superpositions among flavors leading to a rapid growth of entanglement. We notice the almost identical behavior of the $\nu_\mu$ and $\nu_\tau$ initial states in Fig.~\ref{Tang 10GeV}b,c reflecting the close-to-maximal "atmospheric" $\nu_\mu/\nu_\tau$ mixing.

As has been shown already for the concurrence in \cite{Blasone:2015lya}, for larger oscillation baselines and similar to the behavior of oscillation probabilities, the oscillatory pattern of the tangle gets damped, and converges to a constant value in the large $L$ limit. As we will see, this effect can be understood as a consequence of wave packet separation (the decoherence limit). Interestingly, the tangle doesn't 
converge to zero,
implying that some entanglement can survive even over astronomical distances. The graphs also differ significantly from entanglement
measures in the plane wave approach in so far as the tangles don't oscillate back to zero, i.e., the unentangled product state, after a full oscillation cycle, as for example in \cite{Blasone:2007vw,Jha:2020dav}. 
It thus can be concluded that it is the wave packet separation effect that makes the mode entanglement irreversible. Also shown in Fig.~\ref{Tang 10GeV} are the corresponding tangles 
for maximal CP violation, i.e. $\delta = \pi/2$. As canbe seen, 
a non-vanishing Dirac CP phase affects the interference patterns, leading to modifications in the entanglement behavior.

\begin{figure}[h!]
    \centering
    \begin{subfigure}[b]{0.6\textwidth}
        \includegraphics[width=\linewidth]{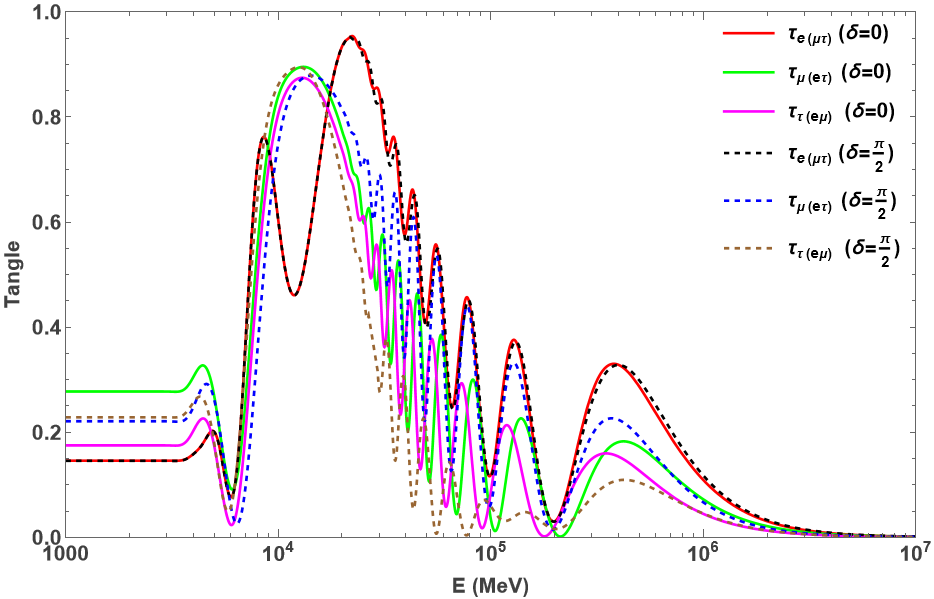}
        \caption{e-channel}
        \label{Tangs2edep}
    \end{subfigure}
    \begin{subfigure}[b]{0.6\textwidth}
        \includegraphics[width=\linewidth]{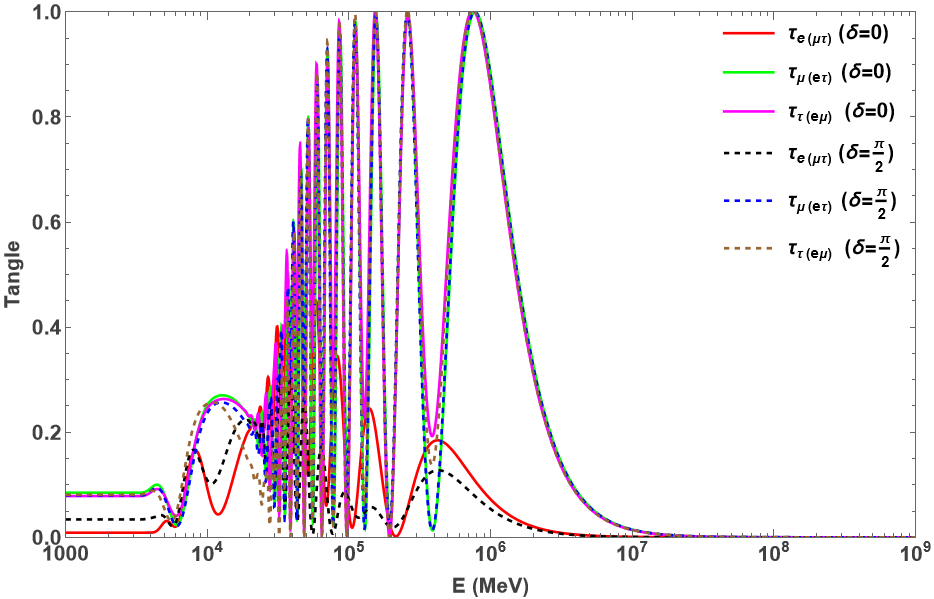}
        \caption{$\mu$-channel}
        \label{Tangss2edep}
    \end{subfigure}
    \hfill
    \begin{subfigure}[b]{0.6\textwidth}
        \includegraphics[width=\linewidth]{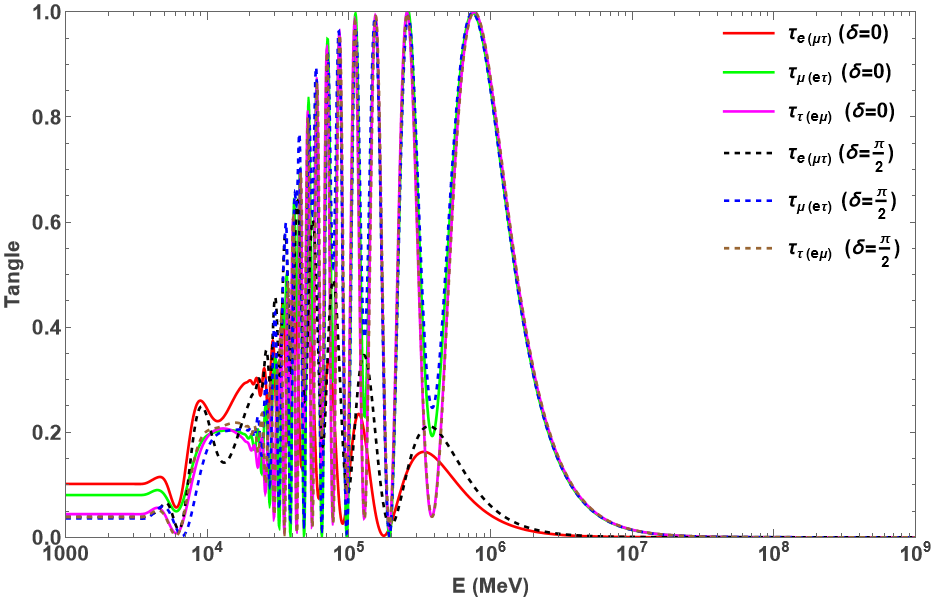}
        \caption{$\tau$ channel}
        \label{Tang32edep}
    \end{subfigure}
    \caption{As Fig.~\ref{Tang 10GeV}, 
    as a function of energy $E$ (in MeV), with baseline $L$ chosen as $10^{15}$ meters.}. 
    \label{Tang Edep}
\end{figure}

Figure~\ref{Tang Edep} depicts the tangle shown in Fig.~\ref{Tang 10GeV} as a function of energy. The respective tangles adopt a similar oscillatory pattern as in Fig. 2  as functions of distance, while the high energy limit corresponds to the small distance limit and vice versa, reflecting the 
$L/E$ dependence of the neutrino oscillation process. Thus the entanglement is suppressed at high energies where no oscillations have occurred yet. The decoherence limit with constant entanglement appears now at small energies, thereby showing that this limit is
correctly interpreted as a consequence of wave packet separation
arising at large $L/E$. 

\begin{figure}[h!]
    \centering
    \begin{subfigure}[b]{0.6\textwidth}
        \includegraphics[width=\linewidth]{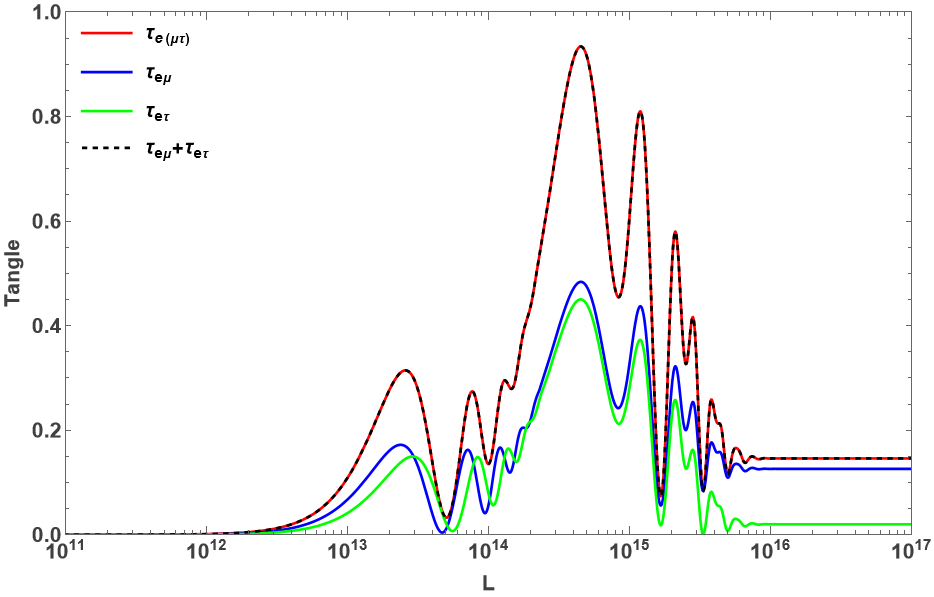}
        \caption{e-channel}
        \label{ckwe}
    \end{subfigure}
    \begin{subfigure}[b]{0.6\textwidth}
        \includegraphics[width=\linewidth]{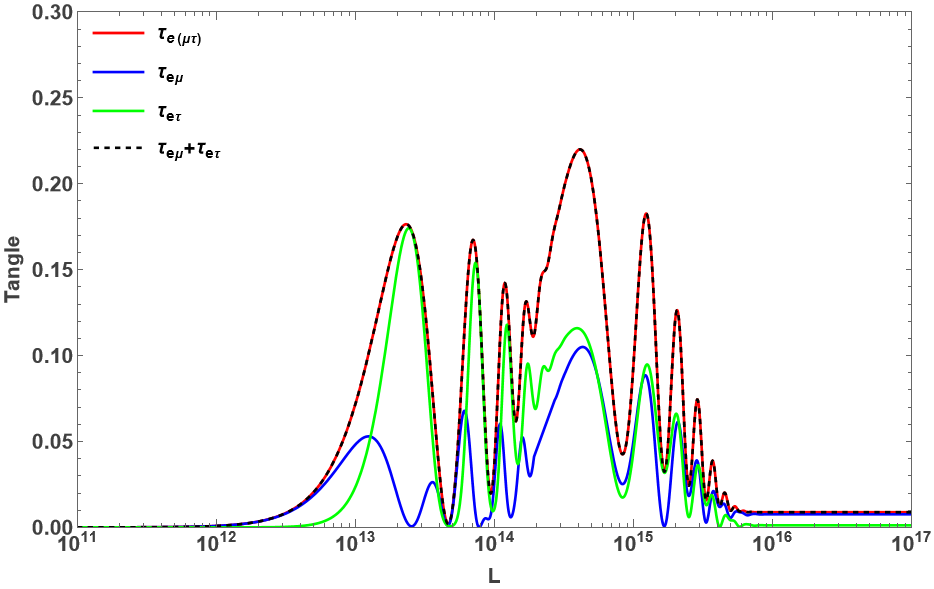}
        
        \caption{$\mu$-channel}
        \label{ckwmu}
    \end{subfigure}
    \hfill
    \begin{subfigure}[b]{0.6\textwidth}
        \includegraphics[width=\linewidth]{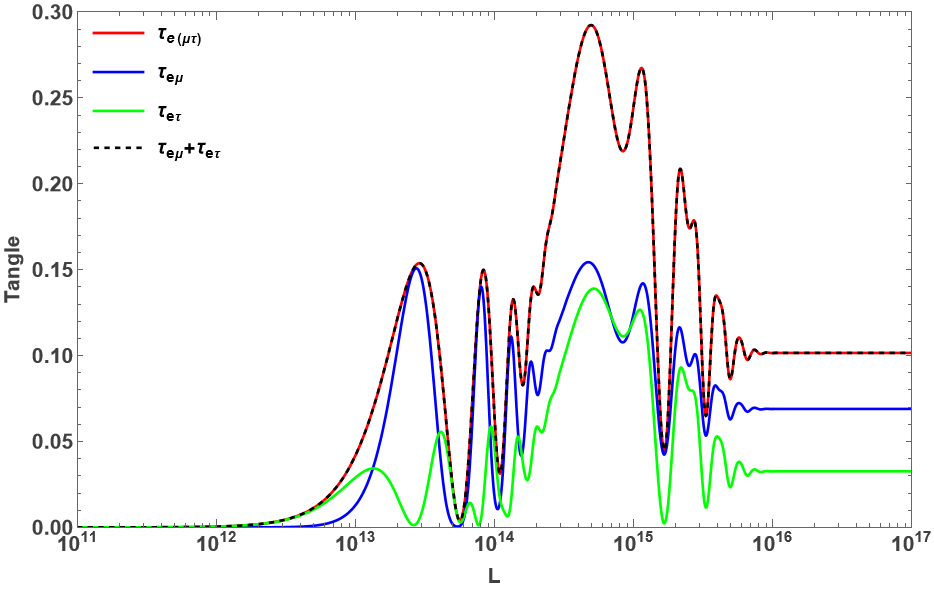}
        \caption{$\tau$ channel}
        \label{ckw tau}
    \end{subfigure}
    \caption{Test of the CKW inequality in terms of tangles, for a $\nu_e$ (a), $\nu_\mu$ (b) and $\nu_\tau$ (c) initial state, respectively, plotted as a function of the oscillation baseline $L$ (in meters). The energy is chosen as $E=10$~GeV and the CP-violating phase is set to zero, $\delta=0$. Note that the CKW inequality is saturated for tangles.} 
    \label{ckw}
\end{figure}

In Fig.~\ref{ckw} we confirm that the distribution of entanglement for an oscillating and decohering neutrino, characterized by the tangles between two individual flavors as well as the tangle between an individual flavor and a pair of flavors, fulfills the  CKW inequality. Note that the tangle between each individual qubit 
and the remaining pairs of qubits is equal to sum of tangles between
the indivudual qubit and one of the qubits in the bipartite state,
implying that the CKW equation is saturated, i.e., it becomes an equality. This implies that the residual entanglement vanishes, i.e., $\tau_e=\tau_{e(\mu\tau)}-\tau_{e\mu}-\tau_{e\tau}=0$. Consequently, the genuine tripartite entanglement of an oscillating neutrino in terms of tangle is not well-defined.
This behavior is typical for genuine tripartite entanglement \cite{Kim:2008juv} of a quantum state belonging to the W class of states.

\begin{figure}[h!]
    \centering
    \begin{subfigure}[b]{0.6\textwidth}
        \includegraphics[width=\linewidth]{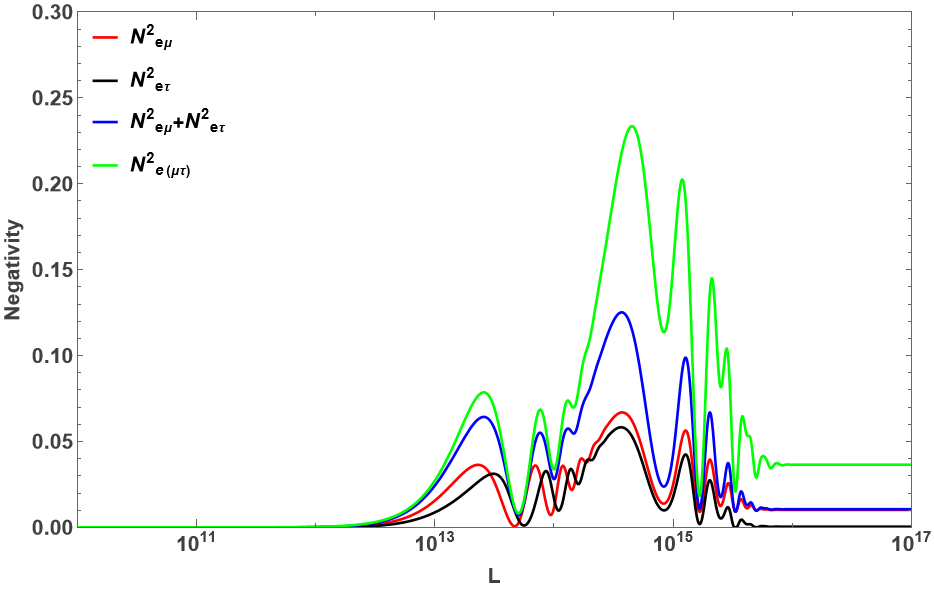}
        \caption{e-channel}
        \label{nega e}
    \end{subfigure}
    \begin{subfigure}[b]{0.6\textwidth}
        \includegraphics[width=\linewidth]{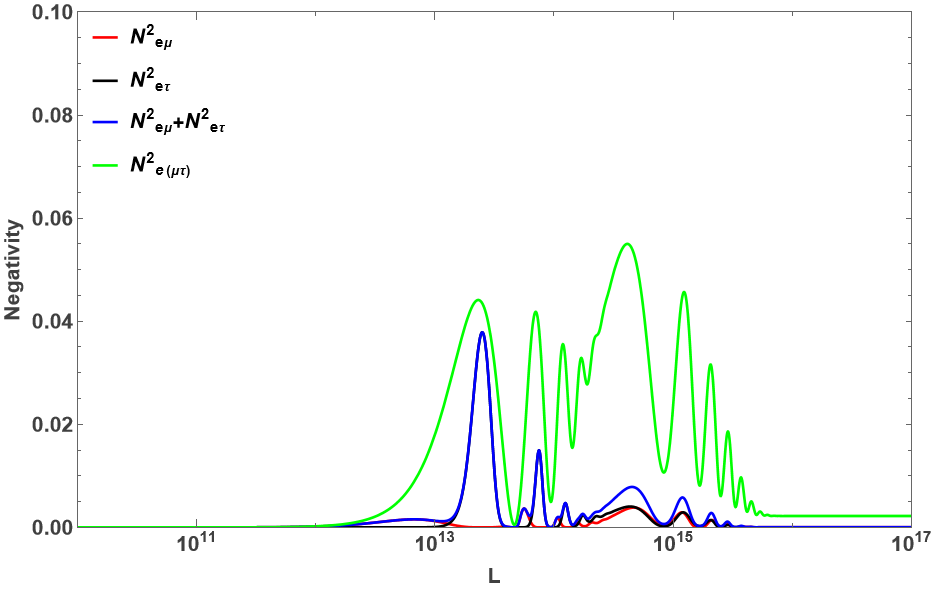}
        \caption{$\mu$-channel}
        \label{nega mu}
    \end{subfigure}
    \hfill
    \begin{subfigure}[b]{0.6\textwidth}
        \includegraphics[width=\linewidth]{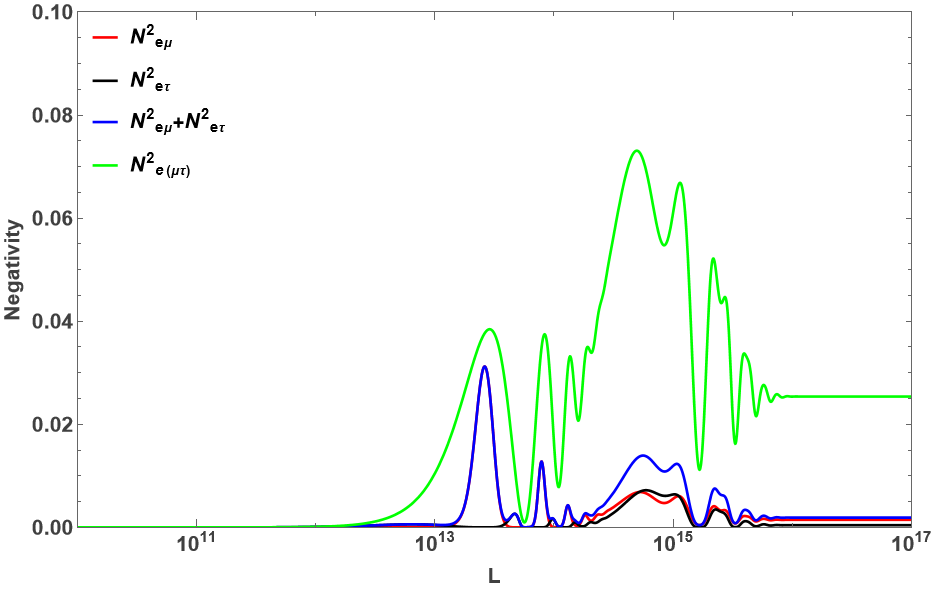}
        \caption{$\tau$-channel}
        \label{nega tau}
    \end{subfigure}
    \caption{Test of the CKW inequality in terms of negativity,
    for a $\nu_e$ (a), $\nu_\mu$ (b) and $\nu_\tau$ (c) initial state, respectively, plotted as a function of the oscillation baseline $L$ (in meters). The energy is chosen as $E=10$~GeV and the CP-violating phase is set to zero, $\delta=0$. Note that in contrast to the case for tangles before, the CKW inequality is not saturated for negativities.} 
    \label{nega}
\end{figure}

The graphs in  Fig. \ref{nega} exhibit the distribution of tripartite quantum entanglement of the neutrino system in terms of negativity.  In all three channels ((a) electron, (b) muon, and (c) tau), an initial rise in negativity is observed, indicating the onset of entanglement generation. As $L$ increases, oscillatory behavior emerges, reflecting the dynamical interplay of quantum coherence and phase evolution. Notably, the green curve, representing the total tripartite negativity $N^2_{e\mu\tau}$, exhibits a dominant behavior at large distances. It suggests the survival of multipartite entanglement at large oscilation baselines for all flavors.
Interestingly, and in contrast to the behavior of tangles, the CKW equation in terms of negativities is not saturated. It means that the CKW equation, which is typically an exact equality when expressed in terms of the tangle  or squared concurrence, reduces to an inequality when other entanglement measures, such as negativity, are used.

\begin{figure}[h!]
    \centering
    \begin{subfigure}[b]{0.6\textwidth}
        \includegraphics[width=\linewidth]{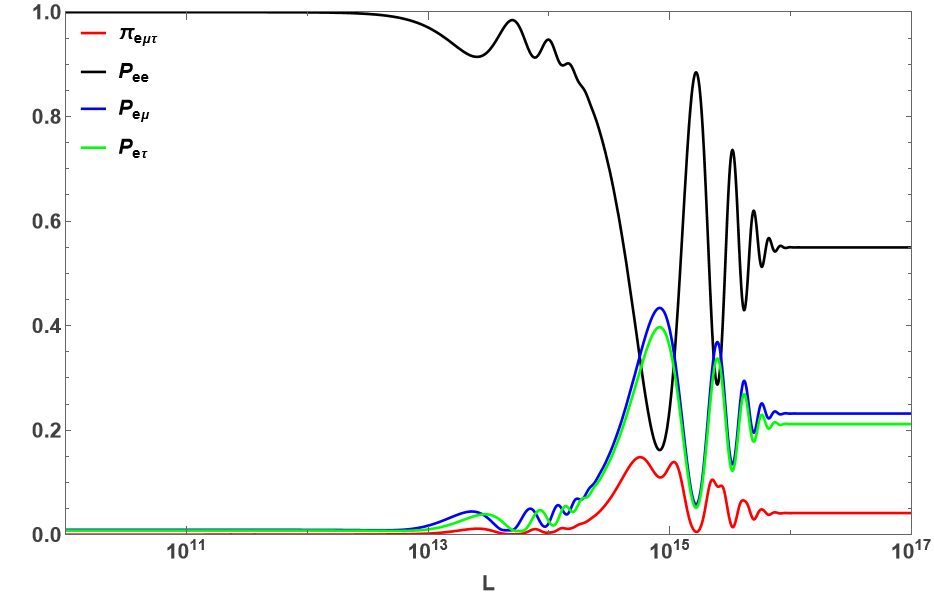}
        \caption{e-channel}
        \label{prob e}
    \end{subfigure}
    \begin{subfigure}[b]{0.6\textwidth}
        \includegraphics[width=\linewidth]{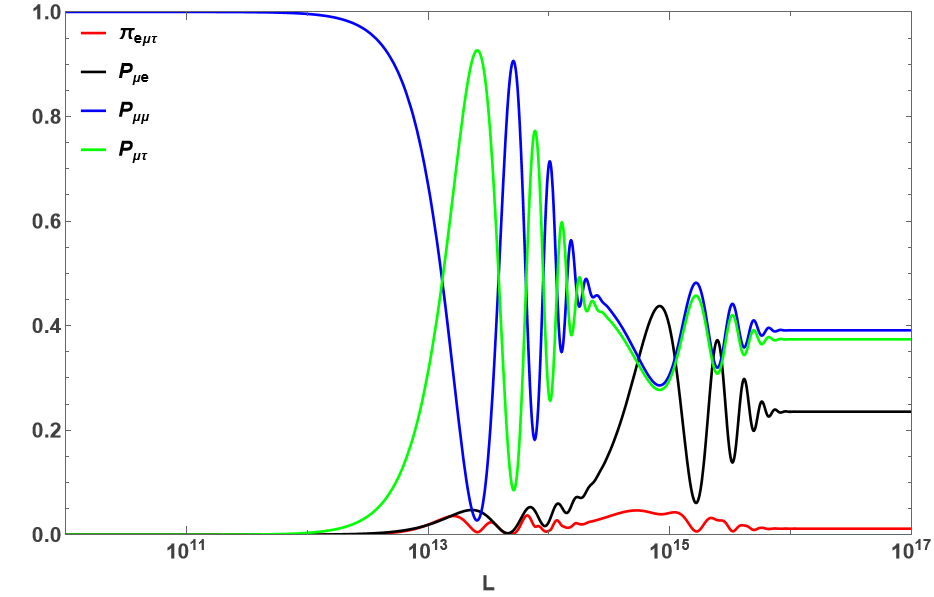}
        \caption{$\mu$-channel}
        \label{prob mu}
    \end{subfigure}
    \hfill
    \begin{subfigure}[b]{0.6\textwidth}
        \includegraphics[width=\linewidth]{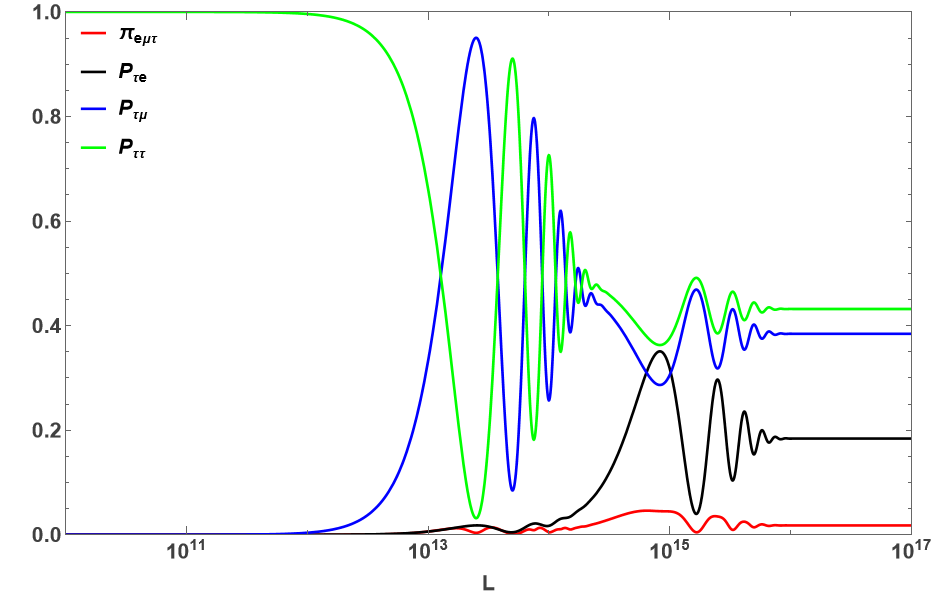}
        \caption{$\tau$-channel}
        \label{prob tau}
    \end{subfigure}
    \caption{The measure of genuine tripartite entanglement, the three-$\pi_{e\mu\tau}$ negativity, compared to the neutrino oscillation probabilities for $\nu_e$ (a), $\nu_\mu$ (b) and $\nu_\tau$ (c) initial states, respectively, plotted as a function of the baseline $L$ (in meters). The energy is set to $E=10$~GeV and the CP-phase $\delta=0$.}
    \label{prob}
\end{figure}

Finally, the graphs in Fig. \ref{prob} illustrate the relationship between the three-$\pi$ negativity measure $\pi_{e\mu\tau}$ and the neutrino oscillation transition probabilities across different channels (electron, muon, and tau) as a function of distance $L$. Fig.~\ref{prob} confirms that an oscillating neutrino features genuine tripartite entanglement and that neutrino flavor oscillations and entanglement are intrinsically linked. The increase of the three-$\pi$ negativity $\pi_{e\mu\tau}$  coincides with the appearance of non-vanishing transition probabilities. It demonstrated that neutrino oscillations are accompanied by an increase in genuine tripartite entanglement and confirms that an oscillating and decohering neutrino can be classified as belonging to the W class of entangled tripartite states. 
This is in fact a direct consequence of analyzing the mode entanglement of a single 
neutrino, implying that only a single non-zero entry can exist in the expressions~(\ref{2.3aa})-(\ref{2.3ac}) in occupation number bassis. 
It can further be seen that genuine, tripartite entanglement persists in the limit of large oscillation baselines, indicating that entanglement in neutrino oscillations is
a macroscopic quantum phenomenon that survives over astrophysical distances, for example, in the case of astrophysical neutrinos studied at neutrino telescopes. 

\section{Conclusions}

Neutrino oscillations provide a unique probe to explore quantum entanglement and coherence over astrophysical distances. In this study, we have analyzed genuine tripartite mode entanglement of an oscillating neutrino in the three-flavor wave packet approach.

We have confirmed the findings of \cite{Blasone:2015lya}, that the tangles, concurrences and logarithmic negativities exhibit an 
oscillatory behavior that does not, however, follow the oscillation probabilities, and that, just like the oscillation probabilities, the entanglement measures converge to a constant value at large oscillation baselines, as a consequence of the separation of wave packets (the decoherence limit). The separation of wave packets also 
implies that the oscillating neutrino does not oscillate back into
the original product state, i.e. it causes entanglement to become
irreversible.

We have confirmed these features for both large $L$ as well as for small $E$, thereby showing that this effect is correctly interpreted as the limit of wave packet separation arising at large $L/E$. Moreover, we have found that concurrence, tangle, and negativity of the tripartite neutrino system characterize the neutrino entanglement structure differently. In particular, concurrence and tangle tend to emphasize the pairwise sharing entanglement and saturate the CKW bound, whereas the negativity is far from saturation. It indicates the presence of non-vanishing residual entanglement and genuine tripartite flavor entanglement. This behavior is further quantified through the three-$\pi$ negativity, which measures the residual tripartite contribution. We also observe that the negativity is always smaller than concurrence and tangle, and it is as expected for a mixed state. These entanglement measures  are found to be sensitive to the values of the Dirac CP phase, resulting in a slight shift in the oscillatory behavior.

We thus can establish within our more accurate and general description the findings of \cite{Jha:2020dav} obtained with the plane wave approximation that three-flavor neutrino oscillations exhibit genuine, tripartite entanglement and that the oscillating neutrino state can be characterized as a member of the W class of entangled states. We also show that these features survive in the decoherence limit.

The results show that genuine multipartite entanglement is a fundamental feature of neutrino evolution and that neutrino entanglement evolves dynamically, with an oscillatory behavior influenced by energy, propagation distance, and CP-violating phases. Decoherence effects lead to a damping of the oscillatory behavior of the entanglement distribution, yet residual entanglement remains, which indicates that quantum coherence is not entirely lost. Thus, 
despite the occurrence of decoherence, neutrinos maintain mode entanglement over large, and even  cosmological distances, in the case of astrophysical neutrinos detected at neutrino telescopes, thereby qualifying neutrinos as an extremely macroscopic quantum phenomenon and a prime probe of the quantum-to-classical transition and potential new effects that may arise e.g. in quantum gravity.

\bibliography{apssamp}



\end{document}